\documentclass{ws-ijmpa}

\begin{document}
\markboth{Daekyoung Kang \& Jungil Lee}
{Inclusive Production of Four Charm Hadrons
at $B$-Factories}

%
\catchline{}{}{}{}{}
%
\title{
INCLUSIVE PRODUCTION OF FOUR CHARM HADRONS \\
AT $B$-FACTORIES
}

\author{Daekyoung Kang and Jungil Lee\footnote{\texttt{jungil@korea.ac.kr}}}

\address{Department of Physics, Korea University, Seoul, 136-701, Korea}

\maketitle


\begin{abstract}
Measurements by the Belle Collaboration of the
exclusive $J/\psi + \eta_c$ and inclusive $J/\psi +c \bar{c}+X$ 
productions in $e^+ e^-$ annihilation
differ substantially from theoretical predictions based on
the nonrelativistic QCD factorization approach.
In order to test if such a discrepancy is originated from the
large perturbative corrections to the hard-scattering amplitude,
we study inclusive production of four charm hadrons
in $e^+e^-$ annihilation at $B$ factories.

\end{abstract}

\vskip 5ex 

The nonrelativistic QCD~(NRQCD) factorization formalism\cite{BBL} 
has enjoyed considerable success in describing production and decay 
rates of heavy quarkonia.
The approach provides an infrared-safe prediction for 
the $P$-wave quarkonium decay\cite{Bodwin:1992ye} 
and explains the large empirical cross sections of prompt 
charmonia at the Fermilab Tevatron.\cite{Braaten:1994vv}
However, there are a few serious challenges to NRQCD.
The first issue is the discrepancy between the NRQCD prediction
and the CDF data for the polarization of prompt $J/\psi$
in large-$p_T$ bins.\cite{PSIPOL} The measurement also confronts
with a recent lattice NRQCD calculation, which supports the dominance of
transverse $J/\psi$.\cite{Bodwin:2004up,Bodwin:2005gg}
The second issue is for the $J/\psi$ production in $e^+e^-$ 
annihilation at $B$-factories.
The cross sections for exclusive production of $J/\psi + \eta_c$
in $e^+e^-$ annihilation measured by the Belle 
Collaboration\cite{Abe:2002rb,Abe:2004ww} 
and by the Babar Collaboration\cite{Aubert:2005tj}
are greater than NRQCD predictions of leading-order 
QCD\cite{LO} by an order of magnitude. 
A few proposals\cite{AFEW} introduced
to solve the problem were disfavored by a recent experimental 
analysis.\cite{Abe:2004ww} The next-to-leading-order QCD corrections 
to the exclusive process have been found to be as large as 
80\%\cite{Zhang:2005ch} of the leading-order prediction.\cite{LO}
It is argued that nonperturbative corrections may be 
large.\cite{NP}
However, these corrections are not yet sufficient to explain the data.
It would be interesting to see if such a large perturbative correction is
also true in inclusive production of four charm hadrons,\cite{Kang:2004mz}
having  a similar parton-level process.

Another large discrepancy between the NRQCD prediction and 
the $B$-factory data
is for the inclusive prompt $J/\psi+c\bar{c}$ production cross section
$\sigma(e^+e^-\to J/\psi +c\bar{c}+X)$. Measured cross 
section\cite{Abe:2002rb} is significantly larger than the NRQCD 
predictions.\cite{Cho:1996cg,Yuan:1996ep,Baek:1998yf}
In order to find if an alternative way could resolve the problem,
a process\cite{Kang:2004zj} was recently studied
within the color-evaporation 
model~(CEM).\cite{Fritzsch:1977ay,%
Halzen:1977rs,Gluck:1977zm,Barger:1979js}
The CEM prediction for the cross section
$\sigma(e^+e^-\to J/\psi +c\bar{c}+X)$ was reportedly 
smaller than the empirical value
by about two orders of magnitude.\cite{Kang:2004zj}
A more comprehensive discussion of the CEM can be found in
Refs.\refcite{Bedjidian:2003gd,Brambilla:2004wf,Bodwin:2005hm}.
The Belle Collaboration also measured the fraction
$R[J/\psi+c\bar{c}]$
of the process within inclusive $J/\psi$ production.
The measured value, $R[J/\psi+c\bar{c}]_{\textrm{Belle}}=%
0.82\pm 0.15\pm 0.14$,\cite{Abe:2002rb,Update} is significantly
larger than the NRQCD predictions of about $15$\%. 
In the CEM, there is only a single universal nonperturbative factor, which 
perfectly cancels in the ratio $R[J/\psi+c\bar{c}]$.
Thus, the CEM prediction for the ratio is expected to be more reliable 
than the absolute values for the cross sections in the numerator 
and the denominator.
Here we briefly review the production of four charm hadrons
in $e^+e^-$ annihilation at $B$ factories.

\vskip 5ex

Since the inclusive four charm hadron production 
involves the same Feynman diagrams for exclusive $J/\psi+\eta_c$ 
production, measuring the cross section
will provide an important information in the estimating the size
of the short-distance coefficient for $J/\psi+\eta_c$ cross section. 
We present $\sigma(e^+e^-\to c\bar{c}c\bar{c}+X)$ prediction in leading
order of strong coupling constant, $\alpha^2\alpha_s^2$.

\begin{figure}
\begin{tabular}{cc}
\psfig{file=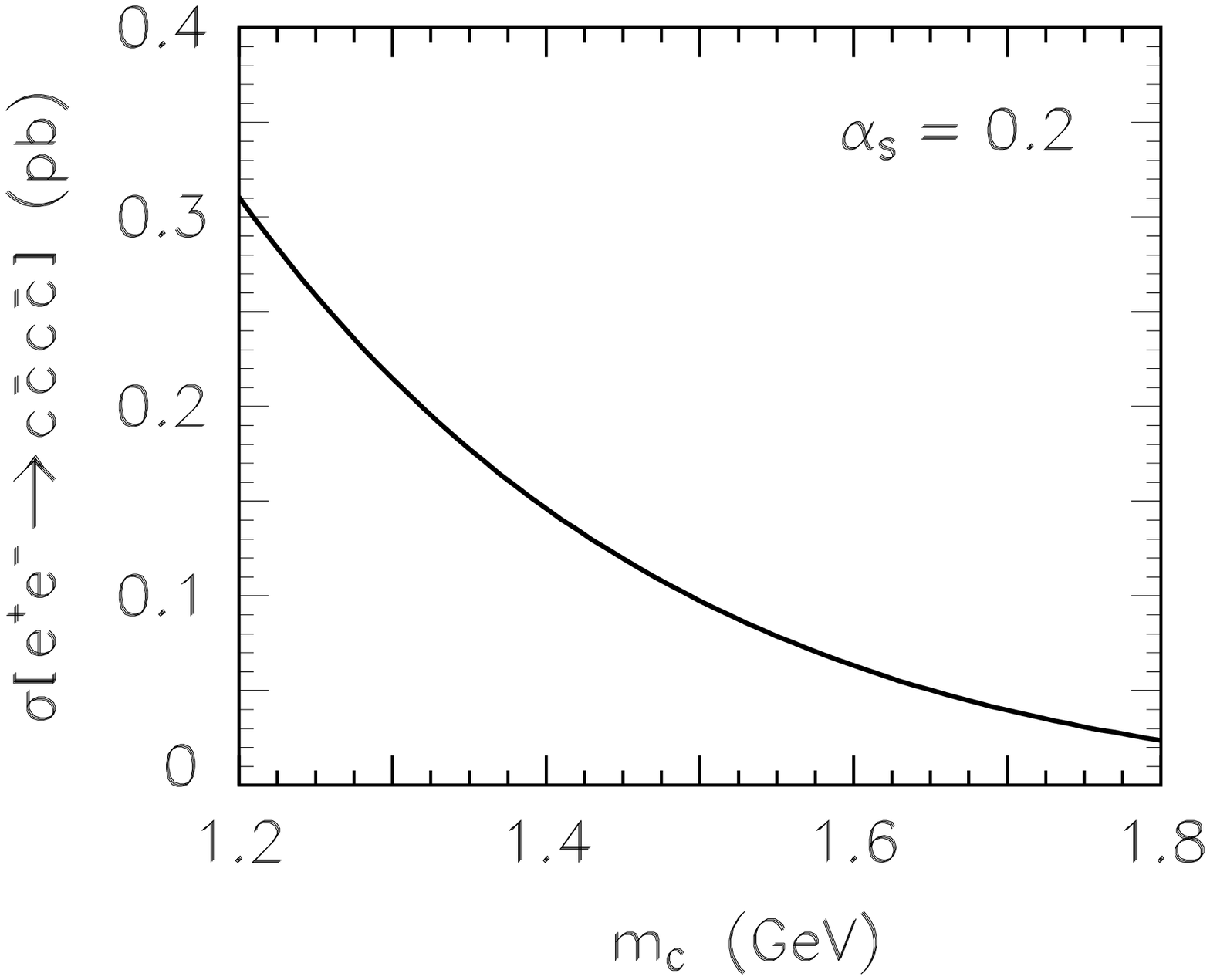,width=6.5cm}&
\psfig{file=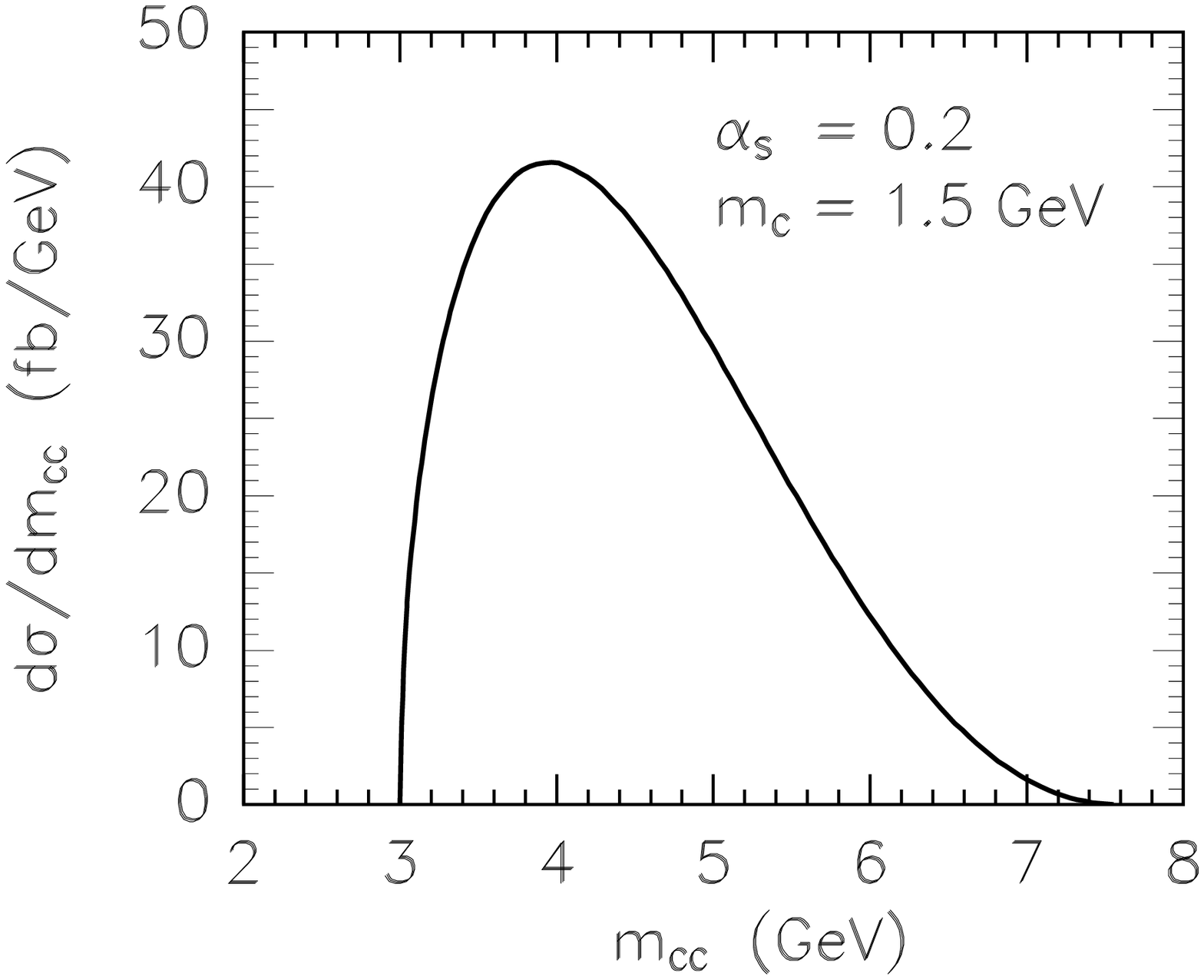,width=6.5cm}\\[-2.0ex]
\textbf{(a)}&\textbf{(b)}
\end{tabular}
\vspace*{8pt}
\caption{\label{fig2} (a)~ 
Total cross section $\sigma(e^+ e^-\to c\bar{c}c\bar{c})$ 
at $\sqrt{s}=10.6$~GeV in pb as a function of $m_c$,
where $\alpha=1/137$ and $\alpha_s=0.2$.
(b)~Differential cross section $d\sigma/dm_{cc}$ in fb/GeV with respect to
the invariant mass $m_{cc}=m_{13}$
of $cc$ for $e^+ e^-$ annihilation into $c\bar{c}c\bar{c}$,
where $m_c=1.5$~GeV,  $\alpha=1/137$, and $\alpha_s=0.2$.
Physical range of the $m_{cc}$ is from
$2m_c$ to $\sqrt{s}-2m_c$.  The area under the curves
are the integrated cross sections 97~fb.
From Ref. 14.
}
\end{figure}
Our predictions for the inclusive four charm hadron cross sections 
in $e^+e^-$ annihilation at $\sqrt{s}=$10.6~GeV depending on the
charm quark mass $m_c$ is  shown in Fig.~\ref{fig2}(a). 
The cross section for 
$e^+e^-\to c\bar{c}c\bar{c}$ is very sensitive to the value of
$m_c$. For $\alpha=1/137$,  $\alpha_s=0.2$, $m_c=$1.5~GeV
$\sigma(e^+e^-\to c\bar{c}c\bar{c})=$97~fb. 
The cross section varies from 
$0.31$~pb at $m_c=$1.2~GeV to $24$~fb at $m_c=$1.8~GeV.
The cross section decreases as $m_c$ increases mainly because 
available phase space shrinks. If one can increase the c.m. energy 
of the $e^+e^-$, 
the $m_c$ dependence will decrease. 

In Fig.~\ref{fig2}(b) we show the differential cross section with respect
to the invariant mass of $cc$. This is the prediction for 
$d\sigma(e^+e^-\to cc+X)/dm_{cc}$ in leading order in $\alpha_s$.
Experimentally, this differential cross section can be compared with
the $\sum_{H,H'}d\sigma(e^+e^-\to HH'+X)/dm_{HH'}$, where $H$ and $H'$
are charm hadrons, which do not include anticharm.

\vskip 5ex

With $\sigma[e^+e^-\to cc+X]\approx 0.1$~pb and
current integrated luminocity $\mathcal{L}\approx 300$~fb$^{-1}$
we expect roughly 30 events will be detected by the Belle detector.
If our leading-order prediction is comparable to the measured value,
it is very probable that the QCD higher-order corrections to the
$J/\psi+\eta_c$ cross section is small. Then the large discrepancy in
$J/\psi+\eta_c$ cross section may be due to the violation of factorization
or existence of new production mechanism. If the measured cross section
for the four charm hadron inclusive production is much larger than
our prediction like the case of $J/\psi+\eta_c$, it is very likely that
perturbative QCD corrections to $J/\psi+\eta_c$ cross section is
large enough to explain the discrepancy, which leads to the failure
of reliability in perturbative expansion.

\section*{Acknowledgments}
We thank Taewon Kim, Pyungwon Ko, and Jong-Wan Lee for their collaboration
on parts of the work presented here.
This work was supported by the Basic Research Program
of the Korea Science and Engineering Foundation (KOSEF)
under grant No. R01-2005-000-10089-0 and by the SK Group 
under a grant for physics research at Korea University.

\end{document}